

\magnification=1200
\baselineskip=20pt
\ \
\def\cl{\centerline}

\def\Mo{M_{\odot}}
\hyphenation{Schwarz-schild}

\vskip 5.0 cm
\cl{\bf VISUAL DISTORTIONS NEAR A NEUTRON STAR AND BLACK HOLE }
\vskip 3 cm
\baselineskip=15pt
\cl {\bf Robert J. Nemiroff }
\bigskip
\bigskip
\bigskip
\cl { NASA / Goddard Space Flight Center }
\cl { Code 668.1 }
\cl { Greenbelt, MD 20771 }

\bigskip\bigskip

\bigskip\bigskip

\cl{ In Press:}

\cl{ AMERICAN JOURNAL OF PHYSICS }

\cl{ 1993, Vol. 61, pp. 619-631}

\vfill\eject

\baselineskip=10pt
\parindent=20pt
\parskip=5pt

\cl {\bf ABSTRACT }

\noindent

The visual distortion effects visible to an observer traveling around and
descending to the surface of an extremely compact star are described.
Specifically, trips to a ``normal" neutron star, a black hole, and an
ultracompact neutron star with extremely high surface gravity, are
described. Concepts such as multiple imaging, red- and blue-shifting,
conservation of surface brightness, the photon sphere, and the existence of
multiple Einstein rings are discussed in terms of what the viewer would
see. Computer generated, general relativistically accurate illustrations
highlighting the distortion effects are presented and discussed. A short
movie (VHS) depicting many of these effects is available to those
interested free of charge.

\vfill\eject

\noindent{\bf I. BACKGROUND }

It is impossible for a human to travel very near a high gravity star which
has a mass like that of the Sun. If, somehow, a person could survive the
extremely harmful radiation that would be emitted on or near these objects,
the high gravity itself would likely pose insurmountable problems.  The
person could not stand casually on the surface of such a star because the
high surface gravity would tend to flatten them.  (Lying down wouldn't
help.)  Were a person to orbit the star in a spaceship, however, the
immense gravitational field might be overcome by a large outward
centrifugal acceleration.$^{1}$ The problem in this case, however, is the
extreme change in gravity between the head and toe of the person, the
extreme tidal pull, would surely prove much more than annoying for any
human.$^{2}$

Nevertheless it is informative and interesting to wonder what it would look
like to visit such a high gravity environment. Significant speculations on
this include popular science fiction stories such as those by Forward$^{3}$
and Niven$^{4}$.   A discussion (with cartoon sketches) of a trip to a
black hole appears in Kaufmann's book ``The Cosmic Frontiers of General
Relativity"$^{2}$.  A description of what hot spots on a high gravity neutron
star would look like to an observer far away is given by Ftaclas, Kearney,
and Pechenick,$^{5}$.  Other descriptions include what a typical neutron
star would look like to a distant observer, including a computer drawn wire
mesh diagram$^{6}$, a description of the sky as seen from the vicinity of a
black hole$^{7-9}$, a description of the image of a thin accretion disk
around a black hole$^{10}$, a description of how the observer would see
self-images near a black hole$^{11}$, and a short computer animated movie
simulating a trip around a black hole while facing the constellation Orion
by Palmer and Unruh.$^{12}$  In general, however, the professional science
literature has focused mainly on mathematical detail rather than observable
image distortions.

In this paper the visual aspects of a journey to several different types of
high gravity stars will be discussed in some detail, along with computer
generated illustrations highlighting the perceived visual distortions.  The
three types of stars that will be discussed are a) a ``normal" neutron
star, b) a black hole, and c) an ``ultracompact" neutron star$^{13}$ having
extremely strong surface gravity. Here the speed of the traveler will
always be considered small when compared to the local speed of light, so
purely special relativistic effects will be ignored.

The paper is structured as follows: Section II discusses the physical
principles and mathematics necessary to describe the perceived visual
distortions. In Section III the types of visual distortions will be
discussed generally. Section IV then proceeds to take the reader on a
fantasy mission to these high gravity environments and describes what
visual distortion effects the viewer would see. In Section V comments are
made.

\bigskip

\noindent{\bf II. GRAVITATIONAL PRINCIPLES AND MATHEMATICS }

The visual distortion that will be described here would be caused by
gravitation in the Schwarzschild metric.$^{14}$  Einstein's general
relativity$^{15}$ is not the only gravitational theory that admits the
Schwarzschild metric as an exterior solution for a spherically symmetric,
non-rotating gravitational field, but it is the preferred theory,
and the theory that will be assumed implicitly here.  The Schwarzschild
metric is
 $$ {\rm ds}^2 = - (1 - R_S/r) c^2 \,  dt^2
                 + (1 - R_S/r)^{-1} \, dr^2
                 + r^2 \, d\theta^2
                 + r^2 {\rm sin}^2 \theta \, d \phi^2 .
 \eqno(1)$$
Here ds is a metric measure of coordinate distance $r$, coordinate time
$t$ and coordinate angles $\theta$ and $\phi$.  The term $R_S$, the
Schwarzschild radius, refers to the radius of a black hole event
horizon, and $c$ refers to the local speed of light.  $R_S$ is directly
proportional to the mass that creates the metric through $R_S = 2GM/c^2$,
where $G$ is the gravitational constant and $M$ is the mass interior to
$r$.

For a photon, ds$^2 = 0$.  Combining this with the conservation of angular
momentum allows one to express the deflection angle $\phi$ of a photon
moving in a gravitational field$^{16}$ as
 $$ \Delta \phi = \int_{r_{emitted}}^{r_{observed}} { dr \over
                  r \, \sqrt{r^2/b^2 - 1 + R_S/r} } ,
 \eqno(2)$$
where $b$ is a constant over the trajectory of the photon path,
corresponding to a linear projected impact parameter of a photon at
infinity for a photon that escapes.  This impact parameter can be
visualized by assuming that when the photon is far from the gravitating
object it travels in a straight line; the impact parameter is the distance
between the closest approach of the continuation of this straight line and
the center of the gravitating object. Note that $\Delta \phi$ is not the
{\it extra} angle deflected by the lens but the {\it total} change in the
$\phi$ angle between the observer and the source, emitted at radial
coordinate $r_{emitted}$ and observed at radial coordinate $r_{observed}$.
This angle is measured with the lens at the vertex, and includes
gravitational deflection.  Therefore, for example, a source seen by an
observer just over the limb of a lens which has only a small mass, and
hence a negligible effect of the trajectory of the photon, has a $\Delta
\phi$ near $\pi$.

An important radius is found from Eq. (2) when $\Delta \phi$ diverges to
infinity. Here a photon will circle the massive star at the photon sphere.
The exact location of the photon sphere is $R_P = 1.5 \, R_S$.  Note that a
``normal" neutron star, one that has the average properties most popularly
attributed to neutron stars in the current scientific literature, does not
have a photon sphere. Were it somewhat more compact, however, it would have
a photon sphere, and were it much more compact, it would have an event
horizon and be called a black hole. For black holes and these
``ultracompact" neutron stars, circular photon orbits can exist.

Photons circling at the photon sphere are not in a stable orbit$^{16}$ -
any small perturbation will cause them to spiral either in or out.  Photons
emitted from infinity with impact parameters slightly greater than $R_B = 3
\sqrt{3} R_S / 2$ will spiral around the compact star near the photon
sphere and then spiral out. Photons emitted from infinity with impact
parameters slightly less than $R_B$ will spiral around near the photon
sphere and then spiral in, eventually colliding with the neutron star
surface or falling into the black hole.  It is also possible for a photon
to be emitted from a ultracompact neutron star surface, orbit near the
photon sphere, and then spiral back in again impacting the surface.  These
describe, in general, all of the distinct cases of photon orbits possible
near a high gravity star.  All shorter photon trajectories will lie on one
of these paths.

Stated differently, the three cases of photon orbits near a gravitating
body can be classified as: ``always outside the photon sphere," ``crossing
the photon sphere," and ``always inside the photon sphere." The first is
the case of a photon passing the neutron star or black hole, reaching a
critical radius $R_c$, and then escaping again toward infinity. In this
case the photon does not reach or cross the photon sphere. Its distance
from the star decreases monotonically until $R_c$, and then increases
monotonically thereafter. The second case is that of a photon continuing to
come toward the neutron star (or black hole) until it impacts the surface
(or falls through the event horizon). Here its distance decreases
monotonically. The third case is that of a photon emitted from the surface
of a ultracompact neutron star, reaching a critical radius $R_c$, and
then falling back down and again impacting the neutron star surface. This
critical radius is given by the cubic equation solution$^{9}$
 $$ R_c = {2 b \over \sqrt{3} } {\rm cos} \,
           \left[ {1 \over 3 } \, {\rm arccos} \,
           \left( { -\sqrt{27} R_S \over 2 b } \right)
           + { 2 n \pi \over 3 } \right] ,
 \eqno(3)$$
where $n = 0$ is for the first case and $n = 2$ is for the third case.

Photons climbing out of a gravitating object become less energetic.  This
loss of energy is known as a ``redshifting", as photons in the visible
spectrum would appear more red.  Similarly, photons falling into a
gravitational field become more energetic and exhibit a blueshifting.
The observed energy $E_{observed}$ at radius $r_{observed}$ of a photon
emitted at radius $r_{emitted}$ with energy $E_{emitted}$ is$^{7}$
 $$ E_{observed} = { \sqrt{ 1 - R_S / r_{emitted} } \over
          \sqrt{ 1 - R_S / r_{observed} } } E_{emitted} .
 \eqno(4)$$
Note that the magnitude of the redshifting (blueshifting) effect is not a
function of the emitted angle or the received angle of the photon - it
depends only on how far radially the photon had to climb out of (fall into)
the potential well.  Also note that the power received from a continuously
emitting source would have additional factors of $\sqrt{1 - R_S/r_{emitted}} /
\sqrt{1 - R_S/r_{observed}}$ caused by the relative differences in the
perceived rate of the number of photons emitted per unit time.

The effect a gravitational field would have on the actual perceived color
of an object is more complex, however, as it depends on the distribution of
photons emitted from the source at different energies relative to the
sensitivity of the observer to measuring photons of different energies. For
example, an object that would be described as green might be very bright in
the ultra-violet - but this would not normally be perceived, as people
cannot see the ultra-violet. Were this object put in a strong gravitational
field and viewed from far away, so that the photons would be significantly
redshifted, the strong ultra-violet emission could be shifted into violet
emission and the object would look more blue, even though its light had
been redshifted. This is an exceptional case, however, and redshifted
objects may indeed appear more red.

\bigskip

\noindent{\bf III. VISUAL DISTORTION EFFECTS IN A HIGH GRAVITY ENVIRONMENT }

\noindent{\bf A. Multiple images and amplification }

A gravitational field may cause a single point source to appear with
multiple images.  For a spherical field all of these images will occur in
the plane defined by the observer position, the center of the lens, and the
source.  These images cannot appear out of this plane because this would
break the principle of conservation of angular momentum along the photon
orbits. Therefore all images of a single point source will appear on a
single great circle on the observer's sky.

A gravitational field may cause an extended source to appear not only
multiply imaged but also greatly distorted. There is at least one feature
that each of the images will maintain, however, that is the same as the
original source: red- or blueshift corrected surface brightness.  Any
radiative process preserves the specific intensity along the beam.$^{17}$
When gravity is involved, power along the beam is not conserved, it grows
or shrinks in accordance with the red- or blueshift.  What is conserved can
be considered to be the ``corrected surface brightness" $B_c = B_r (1 -
R_S/r)^2$, where $B_r$ is the measured surface brightness at $r$. $B_c$
corresponds to the surface brightness of the source measured by an observer
far away from the source, and is considered to be summed over all possible
energies.

For example, if an observer originally saw an unlensed circular source with
constant surface brightness, a gravitational field could cause the observer
to see multiple, elongated, images.  Each image would have, however, the
same surface brightness ($B_c$) as the original unlensed source, after
correction for red- or blueshift factors.

The net flux that reaches the observer from any single image of the source
can be either more or less than the original unlensed flux of the source.
Each image will undergo an amplification $A$, with $A$ not constrained to
be greater than unity.$^{9}$  This means that when considered together, the
images of a source seen near a large gravitational field can have more or
less flux than the same source seen without the intervening gravitational
field. Essentially, there are two types of amplifications a source can be
seen to undergo: time distortion induced amplifications $A_{time}$, related
directly to the slowing of time in a gravitational field that causes
photons to change both their energy (red- or blueshift) and the perceived
arrival rate (and hence the source's perceived power integrated over all
wavelengths), and amplifications in the apparent angular size of the
source, $A_{angular}$.  The total amplification will be designated
$A_{total} = A_{time} \, {\rm x} \, A_{angular}$.  In the convention used
here, all amplifications will be greater than zero.

Time induced amplifications result when the observer is at a different $r$
from the lens center than the source.  When considering only time induced
amplifications, the total bolometric (incorporating all wavelengths) power
received will then be changed by an amount $A_{time} =
(1 - R_S/r_{emitted})^2 / (1 - R_S/r_{observed})^2$.

For the sources near perfect lens - observer alignment, the angular
amplification effects typically dominate over time induced amplification
effects.  Angular amplifications can be computed from the deflection angle
Eq. (2).  If a large change in angular position on the observer's sky
corresponds with a small change in the angular position at the (unlensed)
source location, then the source will appear to be angularly elongated and
hence amplified. Similarly a source can be angularly deamplified, but this
will be referred to as an angular amplification of less than unity. Angular
amplification effects should be computed on the spherical sky of the
observer, and so would be given by $^{8}$
 $$ A_{angular} =  { {\rm sin} \alpha \over {\rm sin} \beta }
                  \left( { {\rm d} \alpha \over {\rm d} \beta } \right)  .
 \eqno(5)$$
Here $\beta$ represents the angular distance between the lens and the
source on the observer's sky in the absence of the gravitational field of
the lens, while $\alpha$ represents this distance in the presence of the
gravitational field and light deflection.  The change in this angular
distance, d$\alpha/$d$\beta$ can be found by application of Eq. (2). When
bending angles due to gravitational effects are small, this reduces to the
amplification formulae given by Refsdal$^{18}$ and Liebes.$^{19}$

The net angular amplification of all the images of a single source can also
be either more or less than the original unlensed flux of the source.$^{7,
20}$  This is because the gravitational field does not change the fact that
the observer still observes the same total angular area as before: $4 \pi$
steradians. Therefore if the apparent angular area from some sources is
greater than without the gravitational field, then there must be other
sources with apparent angular area which is lower, to compensate.  In
practice, only relatively few images from sources that are near the
observer - lens line will have angular amplifications very large
($A_{angular} >> 1$), while the rest of the sources in the sky will be
slightly deamplified ($A_{angular}$ ${{<}\atop{\sim}}$ $1).$

The total flux received by an observer from all the sky can again be either
more or less than the original unlensed flux from all the sky.$^{20}$  A
gravitational field does not create photons - it just redistributes and
(red- or) blueshifts them. The observed angular redistribution and the
relative time distortions, however, now act in opposite directions.  For
the background sky, $A_{angular} < 1$, because now its angular area, which
used to occupy the observer's entire field of view ($4 \pi$ steradians) in
the absence of gravity, is now less by the amount of the angular size of
the photon sphere of the lens.  However, the photons from the sky, because
of the blueshifting, are relatively more energetic and arriving relatively
more often, so that $A_{time} > 1$.$^{7}$

In other words, the background sky takes up less of the observer's sky, but
the observer receives more photons per unit area, and each photon is of
higher energy.   Do these effects exactly cancel?  No. It turns out that
all observers will measure $A_{total} > 1$, with the closer the observer the
greater is $A_{total}$.

\bigskip

\noindent{\bf B. Einstein rings }

An important observational aspect of visual distortions in a high gravity
environment that is discussed more usually in the gravitational lensing
literature than in the introductory gravitation texts is called an Einstein
ring.$^{18-19, 21}$  Before it was shown that all images must occur in the
plane defined by the observer's position, the center of the lens, and the
point source.  But what if these are all collinear?  No plane is then
defined. In this case the image of the point source would appear to the
observer as an infinitesimally thin ring.  This is an Einstein ring. As will
be explained, numerous Einstein rings may appear simultaneously, however,
and they are also important as invisible dividing lines between sets of
images,$^{7}$ even when no source is distorted into a ring.

It is not generally appreciated that the mathematical formalism allows an
infinite number of Einstein rings.  In fact, there can be an infinite
number of Einstein rings for each set of collinear observer, lens, and
source points.  (The infinite nature of these results breaks down as the
assumptions behind the mathematical formalism become invalid.) The only
Einstein ring currently discussed in the literature is the most prominent
one that occurs at precise observer - lens - source alignment, and where
$\Delta \phi = \pi$.  Here light emitted at a specific angle from the
source would be slightly deflected by the gravitational field of the lens
to reach the observer.  Were the source light emitted at a different angle
the lens would either not be able to bend the light enough to reach the
observer or too much.  Since the exact observer, lens, source alignment is
symmetric about the line connecting them, this source would be seen as an
annular ring.  This ring will be referred to as the {\it first} Einstein
ring. (Later the term Einstein ring will be even additionally labelled by
the relative radius of the source.)

Other Einstein rings can be seen angularly closer to the center of the
lens. Photons from the {\it third} Einstein ring (the {\it second} Einstein
ring will be defined two paragraphs below) have fully circled the lens once
near the photon sphere before coming to the observer. In fact, the path of
these photons crosses itself. It is possible for photons to orbit the lens
an arbitrarily large number of times before coming to the observer, and
each of these orbits corresponds to an Einstein ring. Therefore there are
innumerable Einstein rings for this specific observer - lens - source
configuration.  Each Einstein ring is seen successively closer to the
apparent photon sphere position. The more times the photon must circle the
neutron star or black hole before reaching the observer, the more precise
the direction of its emission must have been emitted to attain this
trajectory, the less likely any photon will take this trajectory, the
``dimmer" the Einstein ring.  For this reason the higher order Einstein
rings will usually carry little light when compared to the lower order
Einstein rings. In fact, the relative brightness of each Einstein ring
decreases exponentially.$^{10}$

The first Einstein ring can be seen not only in a high gravity environment,
but also in a low gravity environment quite a distance from much larger
objects, such as normal stars, galaxies, and clusters of galaxies. In fact,
complete first Einstein rings have actually been seen for radio
galaxies.$^{22}$  A good review of extragalactic measurements of
gravitational lens effects is given by Blandford and Narayan.$^{23}$ A good
general review of low gravity gravitational lens effects is given in a book
by Schneider, Ehlers, and Falco.$^{24}$

Another set of Einstein rings is observable when the observer and source
are on the same side of the lens. Then, for compact sources such as a black
hole or a sufficiently compact neutron star, light from behind the observer
is able to make a ``U-turn" around the neutron star and come back to be
visible to the observer.  The Einstein ring seen from these light
trajectories will be called the {\it second} Einstein ring, since it is
seen between the first and third Einstein rings, and is brighter than the
third but dimmer than the first. The fourth Einstein ring in this set is
created when light does a ``U-turn" near the photon sphere of the lens,
then goes all the way around the lens again near the photon sphere, and
finally comes to the observer. Note that there is a critical minimum (or
maximum for observers inside the photon sphere) distance for the photon
just like in the case of slight deflection, that is given by Eq. (3). There
are also an infinite number of higher order Einstein rings of this type.
As before, however, these Einstein rings carry relatively little power when
compared to the lower order Einstein rings.

It is convenient to also define the zeroth Einstein ring, where light from
a source located on the line from the lens through the observer comes
directly undeflected to the observer along a radial line ($\Delta \phi =
0$). This Einstein ``ring" is actually a single point on the observer's
sky.  It differs from the other Einstein rings in that its angular
amplification (of a collinear point source) is not formally divergent.

Note that a single source located precisely on the opposite side of the
lens from the observer would create only the first, third, fifth, etc.
(i.e. odd numbered) Einstein rings.  A single source located on the same
side of the lens as the observer would create the zeroth, second, fourth,
etc. (i.e. even numbered) Einstein rings.

In general, the position of each set of Einstein rings will be different
for each specific source radius from the lens, relative to the observer
position. For example, a point source at infinity directly behind the lens
from the observer would create a complete set of odd numbered Einstein
rings.  A point source located a small, finite distance from the lens (but
still directly behind the lens) would create a different set of odd
numbered Einstein rings.  Each set of Einstein rings can thus be labeled by
the location of the source sphere. Sources at infinity will be referred to
``sky" Einstein rings.  For sources on the surface of the lens, the term
``surface" Einstein rings will be used.  In general, the convention will be
taken of labelling each Einstein ring by the name or radius of the source
sphere.

Mathematically, an Einstein ring will always occur when the total
deflection angle due to gravitation $\Delta \phi$ (Eq. 2) is equal to any
integer multiple of $\pi$ radians.$^{7}$  Note that the Einstein rings are
theoretical constructs and would only be visible were a source placed
precisely on the observer-lens line, which for any small source is
unlikely.

If the angular radius of an opaque lens is larger than the angular radius
of the first Einstein ring for the source, then this ring will not exist in
the sense that it will not appear on the observer's sky.  If the radius of
the lens is smaller than the radius of the first Einstein ring but larger
than the other Einstein rings, then only the first Einstein ring will
exist. If the radius of the lens is small enough so that the lens exhibits
a photon sphere, however, an infinite number of Einstein rings exist. This
is because a subsequent Einstein ring exists for each revolution of the
lens a photon orbit can take, and theoretically, since all of these orbits
are contained completely above the photon sphere, it can take an
arbitrarily large number of them.

It should be noted that the existence of an Einstein ring depends on the
relative positions of the lens, observer, and source, while the existence
of the photon sphere or event horizon does not depend on these relative
positions. It is possible for the first sky Einstein ring to exist for a
given observer looking toward a neutron star lens, but as the observer
moves closer to the neutron star the angular size of the surface becomes
larger than the angular size of this Einstein ring. For black holes and
neutron stars compact enough to have a photon sphere, though, the photon
sphere is a real entity - photons do circle there - whether or not an
observer is there to see them.

\bigskip

\noindent{\bf C. Complete sky and surface visibility }

A complete image of the sky is always contained between each two ``sky"
Einstein rings.$^{7}$  Likewise a complete image of the neutron star is
always contained between each two ``surface" Einstein rings. In general, a
single complete image of all the sources on a sphere centered on the lens
is visible between each two consecutive Einstein rings of that sphere.

If the radius of the lens is small enough so that the lens exhibits a
photon sphere, an infinite number of images can be seen of a source, no
matter its location.  One image of the source comes to the observer
relatively undeflected.  This image is between the zeroth and first
Einstein rings and will be referred to as the primary image. A second image
comes around the opposite limb of the lens from the first image, and
therefore will appear to the observer 180$^{\rm o}$ around the face of the
lens from the first image. This secondary image will always be located
between the first and second Einstein rings. A third image comes around the
same limb as the first image and is seen even closer to the apparent
position of the photon sphere.  This image has circled the neutron star or
black hole fully once before reaching the observer, and its location is
always between the second and third Einstein rings.  The photon path for
this image (and all higher order images) crosses itself. A fourth image
occurs closer to but outside of the same limb as the second image, but has
fully circled the lens once in the opposite direction.  There is a
subsequent image for each revolution of the lens a photon orbits takes, and
theoretically it can take an infinite number of them.  In practice, these
multi-revolution images have little power and would be increasingly hard to
see.$^{9}$

Each set of images contained between successive Einstein rings is converted
into ``mirror writing" with respect to the images between the previous two
Einstein rings. For example, if the source was a book, then the book would
be visible with relatively minor distortions in its primary image - between
the zeroth and first book Einstein rings. For the secondary image, between
the first and second book Einstein rings, the book would appear in mirror
writing, but right side up. The mapping of the entire sphere onto the
annular ring between the two book Einstein rings would also cause prominent
distortions.  The third image of the book, between the second and third
book Einstein rings, would appear in normal writing again (neither in
mirror writing nor inverted), but even more distorted because of the
decreased relative angular area between these two book Einstein rings.  A
discussion of the parity of lensed images for the brightest two images of
the point lens (considered here) as well as other gravitational lens types,
can be found in Blandford and Kochanek.$^{25}$

Therefore, for a compact enough neutron star, one can see the whole neutron
star surface.$^{5}$  An observer can see the complete surface of a lens
(exactly once) when the first surface Einstein ring is the same angular
size as the surface of the lens.  (A derivation of the angular size of a
sphere of mass $M$, radius $R_*$ at distance $D$ is given in the Appendix.)
When the second surface Einstein ring has equal angular size to the
apparent angular size of the lens surface, two complete images of the
lens surface are visible.

Any lens which has a first surface Einstein ring is completely incapable of
blocking light from any source.  These objects cannot ``eclipse" anything.
This is why a neutron star in a well separated binary system can never
block the light of its binary companion.

Less stringently, any lens with a first sky Einstein ring is incapable of
blocking light from the background sky. Almost all stars in our galaxy are
thus incapable of blocking light from random superpositions of background
objects.  For example, no supernovae in other galaxies are missed because
they are ``eclipsed" by a random superposition of a foreground star in the
Milky Way Galaxy.  Were such a chance superposition to occur (it is very
unlikely), the supernova would be greatly amplified by the gravitational
field of the intervening star rather than diminished by an ``eclipse"
effect. With respect to distant sources, these stars are easily compact
enough to show a first Einstein ring to a distant observer, and are
therefore incapable of blocking the source's light.

Every star in existence, besides the Sun but including even the nearest
stars, has a first sky Einstein ring with respect to an earth bound
observer.  The small angular size of this Einstein ring is currently below
optical resolution, but, for the nearby stars, not below the angular
resolution of many radio observations.  The gravity of these normal stars
is strong enough to bend the background light around them and cause distant
sources to be visible to the observer. Almost none of the nearby stars,
however, would have a second sky Einstein ring, unless they were a neutron
star or black hole. Were the star compact enough to have a photon sphere
surrounding it, then, theoretically, an infinite number of sky Einstein
rings (and hence sky images) could exist.

\bigskip

\noindent{\bf D. ``Self" Einstein rings: Where to see yourself }

A very interesting set of Einstein rings are the ``self" Einstein rings,
where observers can see themselves.  The most well known of these can be
seen when the observer is located at the photon sphere.  There observers
can simply look along the photon sphere, where light travels in a circle,
and see the backs of their heads!$^{11,26}$  All observers in the presence of
a sufficiently compact lens, however, can see themselves. Here, light can
leave the observer, travel around the lens and return to the observer to be
viewed.  Observers would see themselves as a series of Einstein rings.  The
more times light can circle the lens and return to the observer, the more
``self" images the observer can see.  For a lens compact enough to have a
photon sphere, observers can, theoretically, see themselves in every self
Einstein ring: an infinite number of times.

Amusingly, there is a single case where observers can see only a
single image of themselves - and this is the case that is well known$^{26}$
- when observers are at the photon sphere!  Here all the self Einstein
rings actually merge with the photon sphere to form a single observer
image.

Observers who see themselves would be viewing themselves with high
amplification.  This is because the self images observers would see would
be on or near Einstein rings - which carry the highest amplifications.
Therefore gravity can act as a powerful microscope! When at the photon
sphere observers can microscopically view the backs of their heads, and
when far away observers can microscopically view their own eyes. This is
because the light that returns to the observer has left on a nearly radial
trajectory - and the part of the observer most nearly radial is the
observer's own eye. When close to and inside the photon sphere, observers
can inspect annular rings on their heads (or spacecrafts).

\bigskip

\noindent{\bf IV. JOURNEY TO A HIGH GRAVITY STAR }

A fairly detailed description of the distortion effects a space-traveler
(or camera) would see on a visit to a high gravity star is now possible.
The case that will be described first will be a trip to a ``normal" neutron
star: one with a currently popular equation for the interior structure of
the star.  This star is not dense enough to have an event horizon or a
photon sphere.

The second case that will be described is that of visiting a black hole.
This case is more complex in that many bound and unbound photon
orbits exist near the black hole.  There is, however, a somewhat simpler
aspect to describing this case than the previous one in that one does not
have to track surface feature distortions for a black hole.

The third and last case that will be described is that of visiting a
ultracompact neutron star - one with an extreme equation of state for its
interior structure that allows a mean density so high the star has a photon
sphere. This is the most complicated case of all to describe as it involves
all three types of photon orbits described in \S II as well as requiring a
description of both the sky and surface feature distortions.

To more clearly delineate what the viewer would see, a set of computer
generated figures were created that document the distortion effects in
terms of familiar icons.  In these illustrations, the sky in the background
behind the high gravity star was taken to be the night sky as viewed from
present-day earth.  More specifically, the background sky is taken from the
Bright Star Catalogue,$^{27}$ allowing all stellar images as dim as 5th
magnitude to be seen, and stellar images as dim as 7th magnitude may be
amplified into visibility. In the two cases of neutron stars, a map of the
earth was projected onto the surfaces of the stars and allowed to distort.
These figures are, in many aspects, fully general relativistically correct.
The resolution of the figures is about 3 arcminutes (0.05$^{\rm o}$).

Stellar image brightnesses are shown by the area the stellar image takes on
the plots: the area is directly proportional to the flux the observer would
receive from the image.  It was impossible to change the pixel brightness,
so many of the single pixel images would actually be seen dimmer than shown
in the figures. Stellar images were allowed to get brighter or dimmer by
angular amplification effects, but time induced amplification effects have
been suppressed.

Note that for $A_{angular} > 1$  the stellar image flux would actually be
seen as an increase in angular area of the image, so that the amplified
angular area of the stellar images in the computer generated plots are, in
this sense, realistic. However, the distortions in the amplified images
would not be readily observable, as these background images would be
unresolved by the viewer and hence indistinguishable from point sources.  A
small amplification would not cause the image to be resolved.  Stellar
images will therefore always be depicted as circles, even when they undergo
angular amplification, as these convey best the idea of an unresolved point
sources.

Only the two brightest images of all sources were tracked by the computer
programs used. All stars originally 5th magnitude or brighter are plotted
as secondary images, no matter their magnitude after gravitational
distortion.  Stars originally 5th magnitude are only plotted as primary
images, however, if their final post-lensed magnitude was 5 or brighter.
Higher order images undergoing larger angular amplification could
potentially be visible but one would need significantly better angular
resolution so see them (the only exception to this will be Fig. 2p), so
they will be suppressed.  An angular amplification limit of a factor of 100
was placed on all images for plotting purposes.

The hypothetical camera used in the simulations is somewhat fanciful but
has several defining characteristics.  First of all the camera is
asymptotically small so that no general relativistic light bending effects
are important over the length of the camera.  The camera's field of view is
90$^o$ across the middle of the picture.  Lastly, the illustrations that
follow have been ``flat-fielded" so that angular area on the spherical sky
is directly proportional to spatial area on the flat page.

\bigskip

\noindent {\bf A. Journey to a normal neutron star }

This section will describe a trip to a fairly ``normal" sized neutron star.
This neutron star has a hard equation of state$^{28}$ for its internal
matter, the result of which is that the matter in the star is not
compressed enough to have either a photon sphere or an event horizon. The
star is considered here to be non-rotating so that gravity external to its
surface is described by the Schwarzschild metric and the
analysis given in Section II.

It is not necessary to specify a mass of the star for a description of a
trip to it.  All distances can be given in terms of the Schwarzschild
radius of the star, and hence are all scalable by this factor.  Distances
will therefore be given first in terms of the number of Schwarzschild
radii, and second, in parenthesis, in terms of kilometers for a specific
model.  For each star, the specific model used will be for a star of mass
$1.4 \Mo$.  By the simple formula $R_S = 2GM/c^2 = (3 \,\, {\rm km}) \,
M/\Mo$, this mass corresponds to a Schwarzschild radius of 4.2 km.  The
surface of this neutron star will be $R_* = 3 \, R_S = $ 12.6 km.

As the voyage starts, the viewer is moving through space toward the
constellation Orion and the neutron star. At a distance from the neutron
star of about 1000 $R_S$ (4200 km), the neutron star destination comes into
view.  It is first noticeable as a very small fuzzy patch, as depicted in
center of Fig. 1a. At 100 $R_S$ (420 km) from the neutron star as
depicted in Fig. 1b, the neutron star itself becomes visible and the
fuzziness around it becomes resolved into a large conglomeration of
individual secondary stellar images. Here for the first time the viewer can
see detail inside the first sky Einstein ring. (To reiterate, Einstein
rings are themselves usually invisible - they can be thought of as
imaginary dividing lines between image sets.) This image conglomeration
will have the same average surface brightness as the rest of the sky: if
one could distribute the starlight in the rest of the sky about the whole
sky it would appear to have the same apparent brightness.
Although not apparent on the black and white figures, each star appears
slightly blueshifted compared with its original appearance, while the
surface appears more noticeably redshifted.

As the viewer approaches the neutron star it becomes evident that
stars that would have appeared eclipsed behind the neutron star in the
absence of its gravitational field now appear to have two bright images:
one just outside the first sky Einstein ring and one just inside. This is
depicted in Fig. 1c where the viewer is now 25 $R_S$ (105 km) from the
center of the neutron star, 12.4 $R_S$ (52.08 km) from its surface.

The viewer now descends to only 10 $R_S$ (42 km) from the star's center.
Visual distortions appear as in Fig. 1d.  Many of the stellar images
interior to the first sky Einstein ring can be very clearly resolved. No
other Einstein rings are in the field of view. At this distance from the
neutron star surface features can now be clearly resolved.  Surface
distortions can be highlighted by comparing the figure to a standard globe
of the earth. More than half of the surface is visible here - but not the
entire surface.  The sky appears slightly more blueshifted than before,
while the neutron star surface appears slightly less redshifted.

For clarity, the first sky Einstein ring has been drawn in with a dashed
line in Fig. 1d.  Also labels pointing out both the primary and secondary
image of the belt of Orion are shown, as well as labels pointing out both
the primary and secondary image of Sirius.

The viewer has now stopped at 10 $R_S$ (42 km) and begins a circular orbit
of the star.  Figs. 1e-1j show views at relative orbital angles of 5, 10,
90, 180, 270, and 360 to the original viewer position. Fig. 1j is the same
as Fig. 1d but is included to provide continuity in the presented sequence.
It is particularly illuminating to compare Figs. 1d, 1e, and 1f. There the
viewer can most clearly see the position of the first sky Einstein ring by
the effect of the viewer's slight motion.

Note that although all stars in the sky have at least one image (the
primary image between the zeroth and first sky Einstein ring), not all
stars have two images. This is because the neutron star is not compact
enough to allow a second sky Einstein ring at the present observer
location.  Therefore the whole sky image between the first and second sky
Einstein rings is {\it not} entirely visible.

Background stars that would have appeared blocked by the neutron star in
the absence of high gravity now appear greatly amplified and positioned
near the first sky Einstein ring. These stars now have two bright images
that appear on opposite sides of the face of the neutron star. When the
observer is in orbit about the neutron star, stellar images outside the
first sky Einstein ring still appear to revolve (due to the viewer's
motion) in the same general sense as they would in the absence of gravity.
Now, however, many of them have a secondary image visible inside the first
sky Einstein ring.  These images, by conservation of angular momentum, must
remain in the plane defined by the source, observer, and lens center, but
on the opposite side of the neutron star from the first image.  Therefore
these images, although rotating in the same sense (clockwise or
counter-clockwise) as the first images, appear to counter-rotate around the
neutron star center when compared to the stellar images just across the
first sky Einstein ring from them.

The hypothetical point on the sky directly opposite the observer through
the lens center has been transformed into a series of circles, the most
prominent of which is the frequently discussed first sky Einstein ring. A
stellar image can never be seen to cross a sky Einstein ring.  For example,
a single stellar image cannot move from the first complete sky image to the
second complete sky image.  When a star approaches this point, it will
either pass above or below it. If it passes above, the primary image will
appear to become greatly amplified and pass above the first sky Einstein
ring.  If it passes below, then the primary image will appear to become
greatly amplified and pass below the first sky Einstein ring. Stars that
would have been seen to approach this point in the absence of strong
gravity now have images that are seen to approach the circle in the
presence of strong gravity.

The description of the apparent motions of surface features are quite
similar to those of the sky features.  The first surface Einstein ring is
not in the field of view, however, and so a complete image of the surface
is not in view.  It {\it is} possible, then, for some surface features to
completely disappear from view.  Notice that surface features that appear
near the limb of the star are somewhat distorted.  Surface features will
not appear dimmer near the limb. This is a consequence of the conservation
of surface brightness.

The viewer now begins to land on the neutron star.  Fig. 1k shows the
distortions from 8 $R_S$ (33.6 km) and a viewing angle of 15$^{\rm o}$ from
looking directly at the neutron star.  The next figures is sequence, Figs.
1l, 1m, 1n, 1o, and 1p, show the distortions from distances of 7 $R_S$
(29.4 km), 6 $R_S$ (25.2 km), 5 $R_S$ (21 km), 4 $R_S$ (16.8 km), and 3
$R_S$ (12.6 km) while the viewing angle pans up from the star to angles
30$^{\rm o}$, 45$^{\rm o}$, 60$^{\rm o}$, 75$^{\rm o}$, and 90$^{\rm o}$
respectively.

Fig. 1p depicts the distortions a viewer would see from the neutron star
surface looking directly tangent to the surface (90$^{\rm o}$ from looking
directly at the neutron star). From the surface one can no longer see any
sky Einstein rings and hence no secondary images of background stars.  From
the surface, the sky appears more blueshifted than ever before, while the
surface now appears without redshift.  This is because photons from
the sky fall toward the neutron star further than before, while photons from
the surface now do not have to climb out of a gravity well to reach
the viewer.

\bigskip

\noindent{\bf B. Journey to a black hole }

This section will describe a trip to the most compact star imaginable: a
black hole.  A black hole can be thought of as any star compressed so
greatly that it not only has a photon sphere but also an event horizon.
The black hole discussed here will be considered to be non-rotating so that
gravity external to its event horizon is described by the Schwarzschild
metric and the analysis given in Section II.

The trip described can be to a black hole with any given mass (and hence
Schwarzschild radius).  For the example model given here, a black hole with
a mass of 1.4 $\Mo$ will be used. All radial distances will be given first
in terms of the Schwarzschild radius and later, in parenthesis, in terms of
kilometers for this specific model. One cannot simply measure this distance
with a series of meter sticks, though.  This is because, for one reason,
any meter stick closer than the event horizon could not be seen by an
observer outside the event horizon.  A better way of visualizing radial
distance is to picture orbiting the black hole at a fixed distance,
measuring the circumference of the orbit, and dividing by $2 \pi$.

Far from the black hole an undistorted night sky is visible with a very
small patch of fuzz in the center.  As the viewer nears the black hole the
fuzzy patch becomes discernable as an unusual conglomeration of stellar
images. Fig. 2a depicts the view visible from 1000 $R_S$ (4200 km) away.

Fig. 2b shows the black hole from a distance of 100 $R_S$ (420 km).  From
this distance the viewer begins to notice that no light comes from a
circular patch in the direction of the black hole.  The only light that
could possible come to the viewer from this area would be from the black
hole itself.  Since here it is considered that the black hole emits no
light$^{29}$, this area is dark. The angular size of the filled black
circle is the angular size of the photon sphere of the black hole mass and
can be found from the discussion in the Appendix.

Fig. 2c shows the black hole from a distance of 25 $R_S$ (105 km).  Here
the angular size of the black hole has increased and the secondary images,
which are inside  the first sky Einstein ring, are now quite clearly
discernable.

Fig. 2d shows the black hole from a distance of 10 $R_S$ (42.0 km). Here
the viewer should notice that the placements of stellar images near the
black hole have changed greatly when compared to Fig. 2a. Stars nearest to
behind the black hole from the observer now have two bright images. The
brightest star in the illustration (and the sky: Sirius) can be seen to
have two bright images: the brightest primary image in the field of view on
the lower left and a secondary image 180$^{\rm o}$ across the face of
the black hole from it.  Primary and secondary images can always be matched
up by connecting them with a Great Circle (a line on these figures) through
the center of the black hole.  Sirius is not the only star to have two
distinct images, however. Notice that Betelgeuse and each star in the belt
of Orion also has two bright images.  Sirius and the stars in the belt of
Orion have been labelled in Fig. 2d.  In fact, all bright stars visible
in the field have two bright images.  Some dim stars that previously could
not be seen now have been amplified by the gravitation of the black hole to
exhibit observably bright images.  In Fig. 2d, the first sky Einstein ring
has been drawn in with a dashed line.

The first sky Einstein ring, shown in Fig. 2d, is an invisible
circle centered on the black hole and dividing the first complete set of
images (those angularly furthest from the disk of the black hole which lie
between the zeroth and first Einstein rings) from the second complete set
of images. Each image in the first set is always brighter than the
corresponding image in the second set. The second sky Einstein ring
appears in the conglomeration of stellar images near the apparent photon
sphere position, just outside the photon sphere. A complete image of the
sky can be seen between these two Einstein rings.

Note that typically stellar images get much dimmer as one looks closer to
the apparent photon sphere position, but the average surface brightness of
the sky there remains unchanged.  In other words, if Fig. 2d was spun about
the center of the black hole smearing all the star images into a blur, the
inner regions near the apparent position of the photon sphere would appear
to have the same average brightness as the outer regions near the edge of
the illustration.  This is a consequence of conservation of surface
brightness discussed in \S III A.  Due to the proximity of the black hole,
the sky would appear slightly blueshifted compared to the original colors,
though this isn't evident on the black and white figures.

The viewer now does an orbit around the black hole at the radius of 10
$R_S$ (42 km).  The distortions the viewer would see are shown in Figs. 2d
- 2j.  These figures depict viewing angles for relative angular positions
of 0$^{\rm o}$, 5$^{\rm o}$, 10$^{\rm o}$, 90$^{\rm o}$, 180$^{\rm o}$,
270$^{\rm o}$, and 360$^{\rm o}$ around the orbit.  A complete orbit would
encompass, of course, 360$^{\rm o}$ and so Fig. 2j is the same as Fig. 2d.

Fig. 2e, showing a relative 5$^{\rm o}$ orbital angle compared to Fig. 2d,
has several interesting differences with this figure. Stellar images
nearest the first sky Einstein ring have shifted the most.  These images
represent stars that are closest to directly behind the black hole from the
viewer. These images appear to move with the highest angular speeds. This
is because a small angular (unlensed) step of the star from just to the
left of behind the black hole from the observer to just to the right causes
all of its images to move from one side of the Einstein ring to the other.
Apparent angular speeds have no maximum limit.  If one attributes a
distance to the images they can even appear to exceed the speed of light.
Note that images of the same star still appear 180$^{\rm o}$ across the
face of the black hole from each other, and that the brighter image is
outside the first Einstein ring, while the dimmer image is inside.

Remember, an entire single image of the sky is contained between the zeroth
and first sky Einstein rings. It is therefore impossible for an image to
leave this region - it cannot just ``go" across this ring and end up
between the first and second Einstein rings. Stars (in reality) approaching
the nadir point below the black hole from the viewer (moving slowly) have
images that appear to approach the Einstein ring and get very bright
(moving rapidly), and then subsequently recede from this Einstein ring and
return to their original brightness.

Now the viewer will go even closer to the black hole. Fig. 2k shows the
visible distortions from 3 $R_S$ (12.6 km): at twice the distance of the
photon sphere. Here the viewer is looking 45$^{\rm o}$ away from the black
hole.  Note the greater number of clearly resolved secondary images visible
near the black hole's limb.

The viewer now reaches the photon sphere and looks up from the black hole
to peer directly along the photon sphere. Fig. 2l shows the distortions
from this distance: 1.5 $R_S$ (6.3 km).  The viewer looks north.  The self
Einstein ring where viewers could see the backs of their heads is the
photon sphere horizon line dividing the light captured by the black hole
from the the light coming from the sky: it is a horizontal line across the
middle of the figure.  The first sky Einstein ring would be an invisible
horizontal line about $2/9$ of the way toward the top of the plot above the
photon sphere. Since the viewer's location and orientation do not allow
the whole face of the black hole to be visible, both of the bright images
(the primary and secondary image) of a single star are not visible at the
same time.  Those stellar images highly amplified above the Einstein ring
are different than those that appear highly amplified just below the
Einstein ring. The primary images just above the Einstein ring in one
direction will have their secondary image appear just below the Einstein
ring in the opposite direction.  Due to the extreme proximity of the black
hole, the sky now appears with an even greater blueshift than before: each
photon has about 73 $\%$ more energy than it would in the absence of the
black hole's gravity

The viewer now starts along an orbit at the photon sphere, 1.5 $R_S$ (6.3
km) from the black hole.  The position of the first sky Einstein ring
becomes more evident when comparing Figs. 2l, 2m, and 2n which have
relative orbital angles of 5$^{\rm o}$ and 10$^{\rm o}$.

The viewer now descends and looks directly away from the black hole. Fig.
2o shows the distortions from 1.1 $R_S$ (4.62 km). All of the sky images
are now compressed into a hole in the direction opposite the black hole.

Fig. 2p shows the distortions visible from 1.01 $R_S$ (4.242 km) while
looking directly away from the black hole. The black hole now encompasses
almost the complete observer sky.  The small hole at the top is what
remains visible of the outside universe.  In this hole there could appear,
theoretically, an infinite number of complete images of the outside
universe.  The angular amplification $A_{angular}$ of the vast majority of
these images is, however, much less than unity: they are greatly
deamplified.  Every sky image has almost exactly the same $A_{time}$,
though, which corresponds to very strong blueshift.

Fig. 2p, as shown, is not an accurate depiction of the distortions a viewer
would see from this position.  It is included because part of it is correct
and the part that is not is informative.  The part that is correct is the
depiction of the relative amounts of black hole and background sky that are
visible.  The thin fuzzy annular ring is not realistic, however, as the
program plotted mostly just the positions of the secondary images. Only a
handful of primary sky images are visible as most of them have suffered
large angular deamplifications.  As stated in the introduction to \S IV,
the secondary images are plotted by the program regardless of the amount of
deamplification. The outer radial limit of the dim ring marks the position
of the second sky Einstein ring.  Stellar images that would be seen between
there and the apparent photon sphere limb of the black hole would be even
higher order images.  This is the only figure where these images are
noticeable by their absence. The programs were not set up to track these
higher order images, and so they are not shown.

\bigskip

\noindent{\bf C. Journey to an ultracompact neutron star }

This section will describe a trip to a very compact neutron star. This
neutron star must have an extremely soft equation of state$^{30}$ for its
internal matter, the result of which is that the matter in the star is
compressed enough to exhibit a photon sphere but not compressed enough to
exhibit an event horizon.  The star, called ``ultracompact,"$^{13}$ is
considered here to be non-rotating, so that gravity external to its surface
is described by the Schwarzschild metric and the analysis given in Section
II.

The trip described can be to a ultracompact neutron star with any given mass
(and hence Schwarzschild radius), and therefore, as in the previous
sections, all distances will first be given in terms of the Schwarzschild
radius, and later, in parenthesis, in terms of kilometers for a specific
model.  For the canonical specific model, a star will again be used with
mass 1.4 $\Mo$, but this time with a surface of $R_* = 4/3 \, R_S$, the
minimum allowed without violating the dominant energy condition.$^{31}$
More specifically, this hypothetical star has $R_S = 4.2$ km and $R_* =
5.6$ km.

Any object compact enough to have a photon sphere will always appear to
have the apparent size of its photon sphere.$^{32}$ This is because photons
coming to the viewer from the object's limb must have orbited near the
photon sphere before escaping to be seen. Therefore, one cannot measure the
actual stellar radius of such an object by measuring the angular radius
seen.  Were the star to shrink the angular radius would appear to be the
same.  One cannot even measure a decrease in $R_*$ by noting the change in
the positions of background stellar images, because there will be no
change.  The only change the observer could note is that of increased
apparent distortion of surface features.

The hypothetical journey begins 100 $R_S$ (420 km) from the neutron star.
Fig. 3a shows the distortion effects a viewer would see from this distance.
The distortions of the background sky are precisely the same as in the
black hole description in the previous section. Essentially an undistorted
night sky is visible from the viewer's location with a small patch of
barely resolved fuzz in the center.  The constellation Orion is clearly
visible to the right of center.

As the viewer nears the ultracompact neutron star the fuzzy patch breaks up
to become a conglomeration of surface images and secondary star images.
This is shown in the succession of Figs. 3b, 3c, and 3d. Fig. 3b shows the
neutron star from a distance of 50 $R_S$ (210 km).  Fig. 3c shows the
neutron star from a distance of 25 $R_S$ (105 km), while Fig. 3d shows the
neutron star from a distance of 10 $R_S$ (42 km).

There are several interesting aspects of Fig. 3d.  The positions of
the stellar images are exactly the same as if the lens were a black hole.
The blueshifting of the image colors would also be exactly the same. The
first sky Einstein ring has been drawn in, and the primary and secondary
images of Sirius and the belt of Orion are labelled. The whole surface is
visible, and some portions can even be seen to have a second image just
inside the apparent position of the photon sphere.  The first surface
Einstein ring can be seen to have about 95 percent of the radius of the
apparent position of the photon sphere in Fig. 3d.  A complete image of the
whole surface of the neutron star is visible inside this first surface
Einstein ring. Another `mirror written' and greatly distorted version of
the entire surface would be seen in the annular space between the first and
second surface Einstein rings (nearer the limb). Higher order sets of
images are not shown but would be difficult to see as they would occupy
increasingly thinner annular rings seen increasingly close to the photon
sphere limb.

A surface feature near the limb of the neutron star would have the same
surface brightness as an identical surface feature near the center, as seen
by any observer.  This is again a result of the conservation of surface
brightness and that the surface emission is assumed isotropic.

The viewer now does an orbit around the neutron star at the radius of 10
$R_S$ (42 km).  The distortions the viewer would see are shown in Figs. 3d
- 3j.  These figures depict viewing angles for relative angular positions
of 0$^{\rm o}$, 5$^{\rm o}$, 10$^{\rm o}$, 90$^{\rm o}$, 180$^{\rm o}$,
270$^{\rm o}$, and 360$^{\rm o}$ around the orbit.  A complete orbit would
encompass, of course, 360$^{\rm o}$, and so Fig. 3j is the same as Fig. 3d.

Figs. 3e and 3f, showing a relative 5$^{\rm o}$ and 10$^{\rm o}$ angle
compared to Fig. 3d, shows revealing differences when compared with Fig.
3d.  The comparison allows the reader to discern the first sky Einstein
ring fairly easily. As before, the differences in the stellar positions are
discussed in the previous section on black holes.  Comparison of the changes
in the apparent positions of the surface features are quite similar but
harder to discern in the figures.

The viewer is now taken even closer to the neutron star.  Fig. 3k shows the
distortions from 3 $R_S$ (12.6 km): at twice the height of the photon
sphere. The viewer is still looking directly at the neutron star.

Figs. 3l, 3m, and 3n show distortions from 1.5 $R_S$ (6.3 km), the height
of the photon sphere, as the viewer's inspection angle pans up.  The
relative angles the viewer is looking with respect the direction of the
neutron star are 30$^{\rm o}$, 60$^{\rm o}$, and 90$^{\rm o}$,
respectively.

In Fig. 3n, the viewer now looks directly along the photon sphere.  The
self Einstein ring where viewers could see the backs of their heads is the
photon sphere horizon line dividing the land and the sky.  In exact
accordance with the black hole case (Figs. 2l, 2m, and 2n), Fig. 3n shows
the first sky Einstein ring would be an invisible line about $2/9$ths of
the way toward the top of the plot above the photon sphere.  Similarly, the
first surface Einstein ring could be drawn in as a line just under (about a
line's width below) the photon sphere.

The apparent color of the surface features would now appear less
redshifted. This is because the light no longer has to climb out of so deep
a potential well to reach the observer. The surface would still appear
redshifted to some degree, as the light must climb from the surface to
the photon sphere, but not nearly so much as before.  The sky would appear
more blueshifted than before.  In fact, the sky would appear to have the
same appearance and blueshift as if the observer was the same distance from
a black hole.

The viewer now does an orbit at the height of the photon sphere, as shown
in Figs. 3n - 3s.  Fig. 3o shows the view 5$^{\rm o}$ from the Fig. 3n into
the orbit, while Figs. 3p, 3q, 3r, and 3s are 90$^{\rm o}$, 180$^{\rm o}$,
270$^{\rm o}$, and 360$^{\rm o}$ respectively. As the viewer moves along
the photon sphere, both the stellar images and the surface
images appear to move in peculiar ways. Stars that would have appeared
behind the observer in the absence of large gravitational light deflection
effects now have images that appear in front of the viewer and above the
photon sphere, but below the first sky Einstein ring. A star that
approaches the opposite side of the neutron star from the viewer appears to
approach the first Einstein ring from below, as this secondary image
get brighter.  As the motion of the viewer causes the star to move
relatively from behind to in front of the viewer (crossing close to the
opposite side of the neutron star), the secondary stellar image moves much
faster, reaches it maximum brightness, and moves quickly (below the first
sky Einstein ring) out of the picture. Shortly thereafter, the brighter
primary stellar image above the first sky Einstein ring comes into view on
the opposite side of the plot.  As the viewer moves further around in its
orbit, this primary stellar image rises and dims.

Similarly, surface features actually just behind the viewer are visible
well in front of the viewer just below the photon sphere but above the
first surface Einstein ring.

The viewer now descends to the neutron star surface.  This is shown in
Figs. 3t and 3u.  Fig. 3t shows the surface distortions from a height of
1.4 $R_S$ (5.88 km) and Fig. 3u illustrates the distortions visible from
just above the surface, at a height of 1.33 $R_S$ (5.6 km).

The viewer now appears to be in a slight bowl.  Looking horizontally the
viewer does not see the sky but rather the surface.  Even looking at an
angle slightly up, the viewer would be observing the surface, no matter
which azimuthal direction is observed. This is because of the photon orbits
that are trapped.  Some photon orbits leave the neutron star and fall back,
never reaching infinity.  If the viewer looks along these paths, the viewer
will be looking slightly up and still seeing an image of the surface. In
fact the sky now appears scrunched up to occupy a smaller ``hole" above the
observer.  Now the surface appears to have no redshift at all, while the
sky has its maximum blueshift.

As before, the apparent line dividing land and sky still marks the apparent
position of the photon sphere.  Also, as before, when orbiting at the
photon sphere, the sky Einstein rings are all seen above the apparent
photon sphere position. The first sky Einstein ring is about $1/9$ of
the length of the plot border above the photon sphere.  The many surface
Einstein rings and complete images of the stellar surface are contained
just below the photon sphere, although they together occupy only a thin
sliver below the photon sphere.

In the last sequence, the viewer pans around to see the whole neutron star
surface.  This is shown in Figs. 3v - 3w, showing the surface image
distortions at relative viewing angles of 120$^{\rm o}$, and 240$^{\rm o}$.
The entire surface and sky is visible, but not in a single field of view.
Lastly, Fig. 3x has the viewer looking along the original northward
direction, the same as Fig. 3u.

\bigskip

\noindent{\bf V. COMMENTS }

An interesting coincidence is that the size of the ``ultracompact neutron
star earth" pictured in Figs. 1d, 2d, and 3d is only a factor of a four
smaller than the size the earth would be were it actually compacted to
neutron star or black hole densities.  Were the earth compacted to these
densities, however, it would explode. The earth does not have the
gravitational mass needed to suppress the enormous repulsive force between
nuclei at densities this high.

The type of map projections the Schwarzschild metric creates are different
than any commonly used type of map projection,$^{33}$ and different than
any known type of map projection that the author is aware of.

A video has been produced featuring these lens effects$^{34}$ and is
available free of charge.  To receive a copy of the VHS tape, please write
to the author's address given under the paper's title.  The author will try
to maintain an abundance of copies of the video on the latest popular video
display medium (be it HDTV tape, laser disc, etc.) to service requests even
several years after this article's publication.

\noindent {\bf ACKNOWLEDGEMENTS }

I would like to thank Christ Ftaclas and Kent Wood for initially
stimulating interest in the project and for helpful suggestions.
Additionally I would like to thank Peter Becker, Peter Noerdlinger, Kevin
Rauch, Brad Stuart, Mark Stuckey, John Wallin, and Daryl Yentis for advice
and helpful discussions.  This paper is dedicated to the memory of Ana
Nash.

\noindent{\bf APPENDIX }

What is the apparent angular size of a sphere of mass $M$ (corresponding to
Schwarzschild radius $R_S$), radius $R_*$, at distance $D$?  This extremely
simple and beautiful problem is important to observable aspects of high
gravity environments.

The impact parameter at infinity, here labeled $b$, is a constant along the
trajectory of the photon.$^{28}$  At any radius $r$ from a sphere with mass
$M$ with Schwarzschild $R_S$ on the photon's trajectory, this constant is
equal to
 $$ b = { r \, {\rm sin} \delta \over
          \sqrt{1 - R_S / r } } ,
 \eqno({\rm A}1)$$
where $\delta$ is the angle the photon's velocity makes with the radial
direction.

For a lens large enough not to exhibit a photon sphere $(R_* > 1.5 R_S)$,
the limb of the source will be seen when a photon leaving the source
tangentially grazing its surface, such that $r = R_*$ and $\delta = \pi/2$.
Then
 $$ b = { R_*  \over  \sqrt{1 - R_S/R_*} } .
 \eqno({\rm A}2)$$
This photon will reach the observer with $r = D$ and angle to the center of
the lens of $\delta_{obs}$ such that
 $$ b = { D \, {\rm sin} \delta_{obs} \over
          \sqrt{ 1 - R_S / D }  } .
 \eqno({\rm A}3)$$
Equating Eqs. (A2) and (A3) and solving for $\delta_{obs}$ yields
 $$ \delta_{obs} = {\rm arcsin} \left(
                     { R_* \sqrt{ 1 - R_S/D } \over
                       D \sqrt{ 1 - R_S/R_*} } \right) ,
 \eqno({\rm A}4)$$
which is the apparent size of the lens.

When the lens is compact enough to be surrounded by a photon sphere, one
can simply replace $R_*$ in the above expression with $1.5 R_S$, as these
photons now define the apparent size of the lens.  The result is

 $$\delta_{obs} = {\rm arcsin} \left(
                 { 3 \sqrt{3} R_S \sqrt{1 - R_S/D} \over
                    2 D } \right).
 \eqno({\rm A}5)$$
Be careful to assign the correct quadrant to the result.  If $R_* < 1.5
R_S$, then $\delta_{obs}$ will be greater than $\pi/2$.

\vfill\eject

\noindent{\bf REFERENCES }
\parindent=0pt
\baselineskip=10pt

\bigskip

$^{1}$ Although these arguments don't work for a particle orbiting inside
the photon sphere!  See M. A. Abramowicz and A. R. Prasanna, ``Reversed
Sense of the Outward Direction for Dynamical Effects of Rotation Close to a
Schwarzschild Black Hole," submitted.

$^{2}$ This point is raised in several introductory books on astronomy and
gravitation.  See, for example, W. J. Kaufmann III, {\it The Cosmic
Frontiers of General Relativity} (Little, Brown, and Company, Boston,
1977), pp. 120-150.

$^{3}$ R. L. Forward, {\it Dragon's Egg} (Ballantine, New York, 1980).

$^{4}$ L. Niven, {\it Neutron Star} (Ballantine, New York, 1968), p. 9.

$^{5}$ C. Ftaclas, M. W. Kearney, and Pechenick, K. R., ``Hot Spots on
Neutron Stars. II - The Observer's Sky," Astrophys. J. {\bf 300}, 203-208
(1986).

$^{6}$ H.-P. Nollert, H. Ruder, H. Herold, and U. Kraus, ``The Relativistic
`Looks' of a Neutron Star," Astron. Astrophys. {\bf 208}, 153-156 (1988).

$^{7}$ C. T. Cunningham, `` Optical Appearance of Distant Observers near
and Inside a Schwarzschild Black Hole," Phys. Rev. D. {\bf 12}, 323-328
(1975).

$^{8}$ J. Schastok, M. Soffel, H. Ruder, and M. Schneider, ``Stellar Sky
as Seen From the Vicinity of a Black Hole," Am. J. Phys., {\bf 55},
336-341 (1987).

$^{9}$ H. C. Ohanian, ``The Black Hole as a Gravitational `Lens'," Am. J.
Phys. {\bf 55}, 428-432 (1987).

$^{10}$ J.-P. Luminet, ``Image of a Spherical Black Hole with Thin
Accretion Disk," Astron. Astrophys. {\bf 75}, 228-235 (1979).

$^{11}$ W. M. Stuckey, ``The Schwarzschild Black HOle as a Gravitational
Mirror," Am. J. Phys., submitted (1992).

$^{12}$ L. Palmer and W. Unruh, shown at a Texas Symposium on Relativistic
Astrophysics in the late 1970s.

$^{13}$ B. R. Iyer, C. V. Vishveshwara, S. V. Dhurandhar, ``Ultracompact (R
less than 3 M) Objects in General Relativity," Class. Quant. Grav. {\bf 2},
219-228 (1985).

$^{14}$ K. Schwarzschild, ``Ueber das Gravitationalsfeld einer Massenpunktes
nach der Einsteinschen Theorie," Sitzunsgsber. dtsch. Akad. Wiss. Berlin,
189-196 (1916).

$^{15}$ A. Einstein, `` Die Grundlage der allgemeinen
Relativit\"atstheorie," Ann. Phys. {\bf 49}, 769-822 (1916).

$^{16}$ See, for example, S. Chandrasekhar, ``The Mathematical Theory of
Black Holes," (Clarendon, Oxford, 1983).

$^{17}$ See, for example, G. B. Rybicki and A. P. Lightman, {\it Radiative
Processes in Astrophysics} (Wiley, New York, 1979), p. 7.

$^{18}$ S. Refsdal, ``The Gravitational Lens Effect," Mon. Not. Roy.
Astron. Soc. {\bf 128}, 295-306 (1964).

$^{19}$ S. Liebes Jr., ``Gravitational Lenses," Phys. Rev. B. {\bf 133},
835-844 (1964).

$^{20}$ A discussion for large $r/R_S$ is given by Y. Avni and
I. Shulami, ``Flux Conservation by a Schwarzschild Gravitational Lens,"
Astrophys. J. {\bf 332}, 113-123 (1988).

$^{21}$ A. Einstein, ``Lens-like Action of a Star by the Deviation of Light
in the Gravitational Field," Science {\bf 84}, 506 (1936).

$^{22}$ The discovery paper for the first ``radio ring" is: J. N. Hewitt,
E. L. Turner, D. P. Schneider, B. F. Burke, G. I. Langston, and C. R.
Lawrence, ``Unusual Radio Source MG1131+0456 - A Possible Einstein Ring,"
Nature {\bf 333}, 537-540 (1988).

$^{23}$ R. D. Blandford and R. Narayan, ``Cosmological Applications of
Gravitational Lensing," Ann. Rev. Astron. Astrophys. {\bf 30}, (1992)
in press.

$^{24}$ P. Schneider, J. Ehlers, and E. E. Falco, {\it Gravitational
Lenses} (Springer Verlag, Berlin, 1992).

$^{25}$ R. D. Blandford and C. S. Kochanek, ``Gravitational Lenses," in
{\it Dark Matter in the Universe, Volume 4, Jerusalem Winter School for
Theoretical Physics}, eds: J. Bahcall, T. Piran, and S. Weinberg (World
Scientific, Singapore, 1987), pp. 133 - 205.

$^{26}$ F. H. Shu, {\it The Physical Universe, An Introduction to
Astronomy} (University Science Books, Mill Valley, 1982), pp. 137-138.

$^{27}$ D. Hoffleit, "The Bright Star Catalog, 4th Revised Ed.," (Yale, New
Haven, 1982).

$^{28}$ See discussions in: S. L. Shapiro and S. A. Teukolski, {\it Black
Holes, White Dwarfs, and Neutron Stars} (Wiley, New York, 1983).

$^{29}$ But black holes can emit radiations as they evaporate, for a good
discussion on this, see S. W. Hawking, ``The Quantum Mechanics of Black
Holes," Sci. Am. {\bf 236}, 34-40 (1977).

$^{30}$ There has been some discussion as to whether a neutron star can
exist in a state this compact.  For some discussion on this see C. E.
Rhoades Jr. and R. Ruffini, ``Minimum Mass of a Neutron Star," Phys. Rev.
Lett.  {\bf 32}, 324-327 (1974).  An equation of state for a neutron star
that allows a neutron star to be this compact is given in S. Bahcall, B. W.
Lynn, and S. B. Selipsky ``New Models for Neutron Stars," Astrophys. J.
{\bf 362}, 251-255 (1990). A discussion including the relevant physical
principles involved is given in A. P. Lightman, W. H. Press, R. H. Price,
and S. A. Teukolski, {\it Problem Book in Relativity and Gravitation}
(Princeton, Princeton, 1975).

$^{31}$ For a lively discussion of extreme energy conditions see M. S.
Morris and K. S. Thorne, ``Wormholes in Spacetime and Their Use for
Interstellar Travel: A Tool for Teaching General Relativity," Am. J. Phys.
{\bf 56}, 395-412 (1988).

$^{32}$ J. van Paradijs, ``Possible Observational Constraints on the
Mass-Radius Relation of Neutron Stars," Astrophys. J. {\bf 234}, 609-611
(1979).

$^{33}$ This possibility was pointed out to me by K. Wood in 1990.

$^{34}$ R. J. Nemiroff, ``Trip to a Neutron Star: The Movie," Bull. Am.
Astron. Soc. {\bf 23}, 1418 (1991).

\vfill\eject

\noindent{\bf FIGURE CAPTIONS }
\baselineskip=10pt

\noindent {\bf Fig. 1:} Visual distortions near a ``normal" neutron star.
The neutron star depicted has neither a photon sphere nor an event horizon.

\noindent {\bf Fig. 2:} Visual distortions near a black hole.  A black hole
has both a photon sphere and an event horizon.

\noindent {\bf Fig. 3:} Visual distortions near an ultracompact neutron
star.  This hypothetical neutron star has a photon sphere but no event
horizon.

\vfill\eject

\end